\documentclass[aps,10pt,superscriptaddress,prl,twocolumn]{revtex4}
\usepackage{epsfig,graphics,amssymb,amsmath,subeqnarray,color}
\usepackage[applemac]{inputenc}
\usepackage[final]{pdfpages}
\begin{document}
\title{Geometrically-protected reversibility in hydrodynamic Loschmidt-echo experiments}
\author{Raphaël Jeanneret}
\affiliation{PMMH-CNRS UMR 7636-ESPCI ParisTech-Universit\'e Paris 6-Universit\'e Paris 7, 10, rue Vauquelin, 75005 Paris, France}
\author{Denis Bartolo}
\affiliation{PMMH-CNRS UMR 7636-ESPCI ParisTech-Universit\'e Paris 6-Universit\'e Paris 7, 10, rue Vauquelin, 75005 Paris, France}
\affiliation{Ecole Normale Sup\'erieure de Lyon, CNRS UMR 46 all\'e d'Italie, 69007 Lyon, France }
%
%%%%%%%%%%%%%%%%%%%%%%%%%%%%%%
\begin{abstract} 
We  demonstrate an archetypal Loschmidt-echo experiment involving thousands of  droplets which interact  in a reversible fashion via a viscous fluid. Firstly, we show that,  unlike equilibrium systems,  periodically driven  microfluidic emulsions  self-organize and geometrically protect  their macroscopic reversibility.   
Self-organization is not merely dynamical; we show that it has a clear structural signature akin to that found in a mixture of molecular liquids. Secondly, we show that, above a maximal shaking amplitude,   structural order and reversibility are  lost simultaneously in the form of a first order  non-equilibrium phase transition.   We account for this discontinuous transition in terms of a memory-loss process. 
 Finally, we  suggest potential applications of microfluidic echo as a robust tool to tailor colloidal self-assembly at large scales.
%120 words
\end{abstract}
\maketitle
%
%%%%%%%%%%%%%%%%%%%%%%%%%%%%%%%%%%%%%%%%%%%%%%%%%%%%%%%%%%%%%%%%
%
The echo protocol consists in studying the evolution  of a system after a reversal in its dynamics. 
Both from a theoretical and an applied perspective echo protocols have attracted much interest in fields as diverse as quantum information~\cite{Gorin2006},  medical imaging, high $T_{\rm c}$ superconductors~\cite{Okuma2011}, fluid mechanics~\cite{Aref1984}, granular~\cite{Slotterback2012} and soft matter~\cite{Hebraud1997,Pine2005,Metzger2012,Keim2013}.
For instance, in statistical and non-linear physics,  the echo dynamics of chaotic systems has been an  area of intense fundamental research since the original debate between Loschmidt and Boltzmann about the emergence of macroscopic irreversibility in systems governed by time-reversible laws at the microscopic level~\cite{swendsen2008}. Conversely, in the condensed  matter and mechanics communities, the echo protocols had been predominantly used as effective characterization methods (e.g. neutron and NMR spin echo) until  publication of a seminal set of experiments by Pine and coworkers~\cite{Pine2005,corte2008}.  These experiments consist in shearing periodically  a  concentrated  suspension, in which actual { irreversible}  collisions overcome the { reversible} { hydrodynamic} interactions between the particles~\cite{corte2008,Metzger2010a}. Unexpectedly, it was found that when decreasing the driving amplitude the system self-organizes, and displays a bona fide non-equilibrium second order transition yielding  a macroscopically reversible state. The authors  hence revealed a behavior somehow opposite to Loschmidt's  gedanken experiment, suggested 137 years ago~\cite{swendsen2008,Gorin2006}: macroscopic reversibility emerges from an underlying time-irreversible dynamics.
A surge of theoretical studies then showed that the irreversible-to-reversible transition  belongs to the universality class of directed percolation~\cite{corte2008,Mangan2008,Menon2009}, thereby identifying reversibility as the trapping in an absorbing state.  In addition subsequent experiments showed that this scenario is relevant to a broader class of classical many-body systems including driven vortices in type-II  superconductors~\cite{Okuma2011} and dense granular media~\cite{Slotterback2012}.  
Notably, until now, no structural change has been observed at the irreversible-to-reversible transition, and  this phenomenon has been referred to as a ''random self-organization''.

%the self-organization responsible for macroscopic reversibility  has not been   associated with any structural ordering, and has been  referred to as ''random self-organization''  
%even though this  phenomena has been hitherto investigated from a condensed matter perspective, the very nature of the self-organization required to induce the (ir)reversible transition remains  rather poorly understood~\cite{Frenkel2008}. 
%In addition, the attempt to address this unusual echo process from a chaotic-dynamical-system  perspective have been scarce~\cite{Pine2005,During2009}.

Here, we introduce an archetypal  Loschmidt echo experiment. We take advantage of a  microfluidic  setup in which more than $N\sim 2\times10^5$ interacting droplets evolve according to a {\em time-reversible dynamics at the microscopic level}. This setup makes it possible to probe the emergence of {\em macroscopic irreversibility} from the single-particle to the entire system level as the droplets are driven in a periodic fashion. We first  provide a quantitative definition of the system reversibility, which does not depend on the resolution of the measurement apparatus.  We  then demonstrate  that macroscopic irreversibility arises only above a minimal driving amplitude, in the form of a first order nonequilibrium phase transition.  Conversely, in the small driving regime we demonstrate that structural order emerges in the emulsion and   geometrically-protects  reversible macrostates. This spatial ordering corresponds to the formation of two coexisting liquid-like phases.
%show that the instantaneous conformation of reversible states are phase separated into two liquid-like phases.  
The simultaneous loss of reversibility and translational order at higher drivings is then explained by investigating  how  this many-body system forgets about its trajectory  upon periodic driving.
%
%the high-dimensional phase space of the emulsion is explored upon periodic driving.
 We close our article by discussing practical applications of these fundamental results, with a special emphasis on large-scale colloidal self-assembly. 
\begin{figure*}
\begin{center}
\includegraphics[width=2\columnwidth]{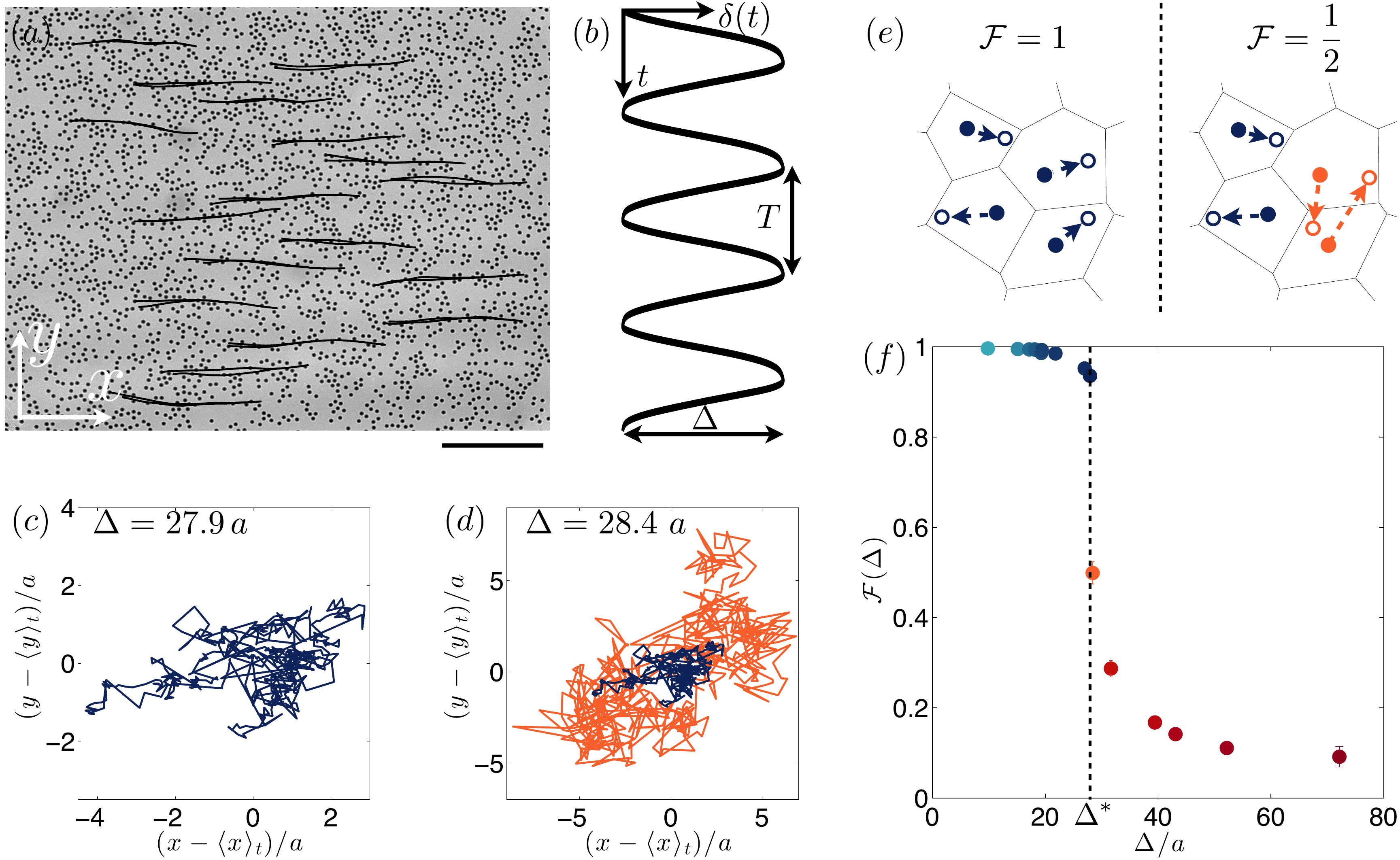}
\end{center}
\caption{(a) Typical snapshot of an experiment. 18 single-cycle trajectories are superimposed  (black lines). (b) Variations of the instantaneous mean displacement $\delta\left(t\right)\equiv\left\langle x_{i}(t)-x_{i}(t=0)\right\rangle_i$ plotted versus time over 4 cycles. $\delta\left(t\right)$ oscillates at a period $T$ with an amplitude $\Delta$. Here, $\Delta=27.9\,a$, and $T=10$ s. Average performed over $\sim 3000$ trajectories. (c)  Strobed trajectory followed by a single particle over 500 cycles (driving amplitude, $\Delta_1=27.9\, a$). (d) Orange curve: Strobed trajectory followed by a single particle over 500 cycles (driving amplitude, $\Delta_2=28.4\, a$). Blue curve: same trajectory as in (c). (e) Schematic picture to guide the definition of the fidelity function ${\cal F}$. The positions of four droplets at the beginning of the cycle (filled circles) define the Voronoï cells. In the left panel, all four droplets  return to their initial Voronoï cells at the end of the cycle (empty circles), which corresponds to ${\cal F}=1$. In the right panel, only two droplets return to their initial  cells (blue circles). The two others escape after a cycle (red circles).  This case corresponds to ${\cal F}=0.5$. (f) Evolution of the fidelity ${\cal F}$ with the driving amplitude $\Delta$. The color of the markers codes for $\Delta$, same color-code throughout the entire document. A threshold amplitude $\Delta^*$ is clearly defined: below $\Delta^*$, ${\cal F}$ remains close to $1$, while above the fidelity decays sharply.}
\label{gouttes}
\end{figure*}

\section*{Results}
\subsection*{A microfluidic echo experiment}
The experimental setup is thoroughly described in the  Methods section. The system is composed of spherical oil droplets dispersed in an aqueous solution, Fig.~1a. This emulsion is confined in a 5 cm-long, and 5 mm-wide microchannel shown in  Supplementary Figure 1. The droplets are highly monodisperse, their diameter $a=25.5 \pm 0.5 \, \mu\rm m$ is comparable to the channel height $h=27 \pm 0.1\, \mu \rm m$. In all that follows the area fraction is set to $\phi=0.36 \pm 0.02$. We use a  high-precision syringe pump to  drive the  oil droplets in a periodic fashion, see Fig.~1a and Supplementary Movie 1. The instantaneous water flow rate is sinusoidal: $Q(t)=Q\sin(2\pi t/T)$, where $T$ is the actuation period. The experiment consists in tracking the  instantaneous position $(x_i(t), y_i(t))$ of $\sim 3000$ droplets at the center of the channel over hundreds of oscillation periods. Already over a single period, the long-range hydrodynamic interactions between the droplets  alter their motion~\cite{Desreumaux}. The trajectories fluctuate around the straight lines parallel to  the $x$-axis that passive tracers would follow, Fig.~1a. However, the center of mass of the particles displays a  well-defined periodic motion along the $x$-direction, with a residual drift one order of magnitude smaller than the smallest driving amplitude. Defining the instantaneous mean particle displacement $\delta\left(t\right)\equiv\left\langle x_{i}(t)-x_{i}(t=0)\right\rangle_i$, we find that $\delta(t)$ oscillates in a sinusoidal manner at a period $T$ with an amplitude $\Delta$ that scales as $QT$, see Fig.~1b.  Within our experimental conditions, $\delta(t)$ deviates by less than $5\,\%$ from an ideal sinusoidal trajectory. We shall also note that the droplets might undergo minute displacements in the $z$-direction (at most  $1.5\,\rm\mu m$). The resulting relative change in their longitudinal velocity would be as small as  $10^{-3}$. In addition, a sole 2D description of the droplet motion has proven to yield excellent agreement with experiments in DC flows~\cite{Desreumaux}. Therefore, we henceforth disregard the droplet fluctuations in the $z$-direction.
 
 $\Delta$ is the sole control parameter of our echo experiments: starting from the same initial conditions, we investigate the global reversibility of the droplet trajectories as $\Delta$ is increased.  We stress that we henceforth focus on the {\em long-time dynamics} of the periodically-driven emulsions, where all the measured quantities have reached their stationary value.
\subsection*{Fidelity decay in shaken emulsions}

Supplementary Movies 2 and 3 show the dynamics of the droplets for two different, but very close,  driving amplitudes $\Delta_1=27.9\, a$, and $\Delta_2=28.4\, a$ respectively. These two movies are strobed at a period $T$.  In Supplementary Movie 2, the particles merely fluctuate around their initial position. It is therefore possible to keep track of each particle by inspecting their strobed pictures only. Increasing $\Delta$ by $2\%$ only yields a markedly different dynamics.  In Supplementary Movie 3, the strobed trajectories are composed of apparently uncorrelated large amplitude jumps, which prohibit tracking the individual-droplet positions. This significant qualitative change in the strobed trajectories suggests that the system undergoes a sharp transition from a reversible to an irreversible state. Let us now provide a quantitative description for this intuitive picture.
The strobed trajectory of a single droplet is shown in Fig.~1c (resp. d) for $\Delta_1$ (resp. $\Delta_2$). Irrespective of the driving amplitude, they look like random walks. Strictly speaking none of the two dynamics is reversible, as even for the smaller amplitude, the strobed trajectory does not amount to a single point. Therefore, the distinction between the two dynamics is intrinsically related to the spatial resolution at which we observe them, see Fig. 1c and 1d. 

In order to minimize the impact of spatial resolution, we quantify the amount of reversibility without referring explicitly to the magnitude of the particle displacements. Here, a  system is defined to be reversible if  the particle positions can be   inferred solely from the inspection of two subsequent strobed pictures. The weakest constraint on the trajectories to solve  this inference problem in a unambiguous fashion is that the particles at $t+T$ do not escape the Voronoï cell they occupied at $t$. Any algorithm based on distance minimization would indeed fail in reconstructing deterministically the droplets' trajectory if this topological condition were not met.
As exemplified in Fig.~1e, we introduce the fidelity function ${\cal F}(\Delta)$ which is the time averaged fraction of droplets that occupy the same Voronoï cell at the beginning and at the end of a period. A perfectly reversible dynamics corresponds to ${\cal F}=1$, whereas a fully irreversible system is characterized by  ${\cal F}=0$. This definition reflects our initial intuitive criteria. We recall here that ${\cal F}$ is a stationary quantity reached after a number of oscillations.
We may note that, not surprisingly, $\cal F$ is an overlap function between two conformations of the system akin  to that defined in~\cite{Abate2007a,Dauchot2010a} to quantify dynamical heterogeneities in amorphous granular ensembles.
%

%%%%%%%%

\subsection*{Irreversibility as a breakdown of self-organization}
\begin{figure*}
\begin{center}
\includegraphics[width=0.7\textwidth]{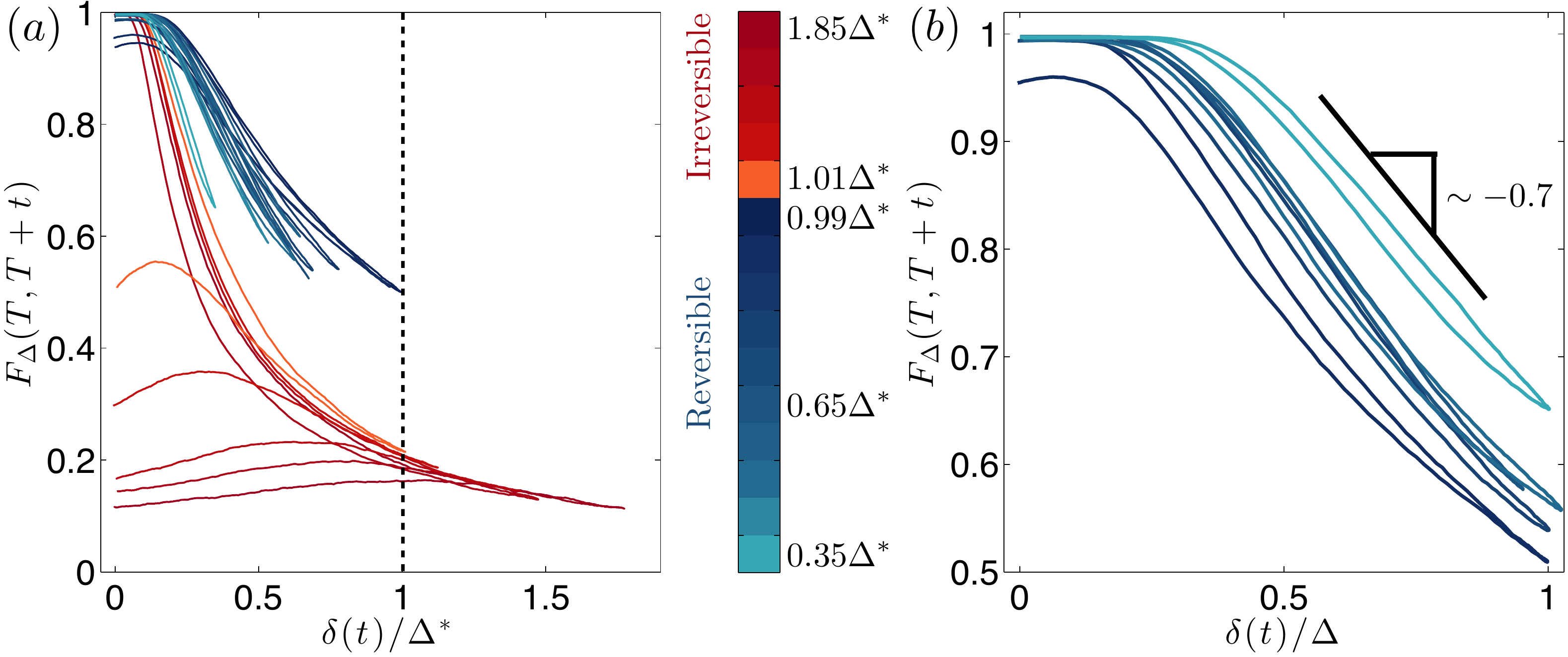}
\end{center}
\caption{(a)  Instantaneous fidelity function $F_\Delta(T,T+t)$ plotted as a function of the rescaled instantaneous displacement  $\delta(t)/\Delta^*$ for all the driving amplitudes. Again, the color codes for the driving amplitude $\Delta$. 
(b) Instantaneous fidelity function $F_\Delta(T,T+t)$ plotted as a function of the rescaled instantaneous displacement  $\delta(t)/\Delta$ in the reversible regime $\Delta<\Delta^*$. For sake of clarity, only half of the experiments below the transition is displayed. The fidelity decays asymptotically as  $\sim-0.7\,\delta(t)/\Delta$, which demonstrates that the emulsion self-organizes differently for each driving amplitude.}
\label{rearrang}
\end{figure*}

 The variations of ${\cal F}(\Delta)$ are plotted in Fig.~1f. ${\cal F}$ remains close to $1$ at small amplitudes. Above a well-identified threshold:  $\Delta^*/a=28.1\pm0.3$, the fidelity decays sharply toward very small values and plateaus as $\Delta$ is further increased. Increasing $\Delta$  by $2\%$ around $\Delta^*$, the number of droplets remaining in their Voronoï cell drops from $94\%$ to $50\%$.  The extreme sharpness of the  fidelity loss is a strong hint of a genuine non-equilibrium phase transition. 
 
Two comments are in order. Firstly, the strong fidelity decay is all the more surprising insofar as the droplets   interact almost exclusively via reciprocal flows. The capillary number, $Ca$, that quantifies how much the droplets are deformed by the flow gradients is  indeed kept below $Ca=10^{-4}$ in all our experiments. Accordingly, we  observe no change in the drop shape in the course of a cycle. The Reynolds number, $Re$, is also much smaller than one both in the oil and in the water phases: $Re<10^{-2}$. In addition, during half a cycle, the fluid momentum diffuses over a distance that scales as $\ell\sim\sqrt{\eta T/\rho}$, where $\eta$ is the water viscosity and  $\rho$ its density. Within our experimental conditions $\ell\sim3$~mm, which is comparable to the channel width. Consequently, the fluid flows in the two phases are very well accounted for by the  Stokes equation which is invariant  under time reparametrization~\cite{guyon2001}. In addition, we stress that the droplets only experience lubricated contacts with the other droplets and the walls. Actual contacts would result in coalescence or wall-wetting events. None of these events were observed in the entire ensembles of $\sim 2\times 10^5$ droplets.  Altogether these four observations imply that  the droplets are chiefly coupled by time-reversible hydrodynamic interactions at all inter-particle distances. We can thus already conclude that any macroscopic fidelity loss must arise from the collective amplification of minute irreversible processes that cannot be probed at the single/two particle levels.

A second important comment is that the uniform flow field in the aqueous phase does not provide any intrinsic length scale  to set the value of $\Delta^*$. This situation contrasts with the previous mechanical echo experiments, which are all based on  a macroscopic shear deformation ~\cite{Pine2005,Slotterback2012,Keim2013}. The non-uniformity of shear flows yields a natural criterion for the "critical" strain amplitude, $\gamma$, at which  particle collisions occur. The fidelity decay is then expected to occur when the relative displacement $\gamma a$ between two particles moving along two parallel flow lines separated by $a$ equals  the typical inter-particle distance $a/\sqrt{\phi}$ (in $2D$). In our Hele-Shaw geometry, the mean flow is a plug flow that can merely advect  the droplet ensemble in a uniform fashion. This observation further supports the hypothesis that the hydrodynamic interactions are responsible for the transition toward a macroscopically (ir)reversible state in our experiments.
\begin{figure*}
\begin{center}
\includegraphics[width=0.7\textwidth]{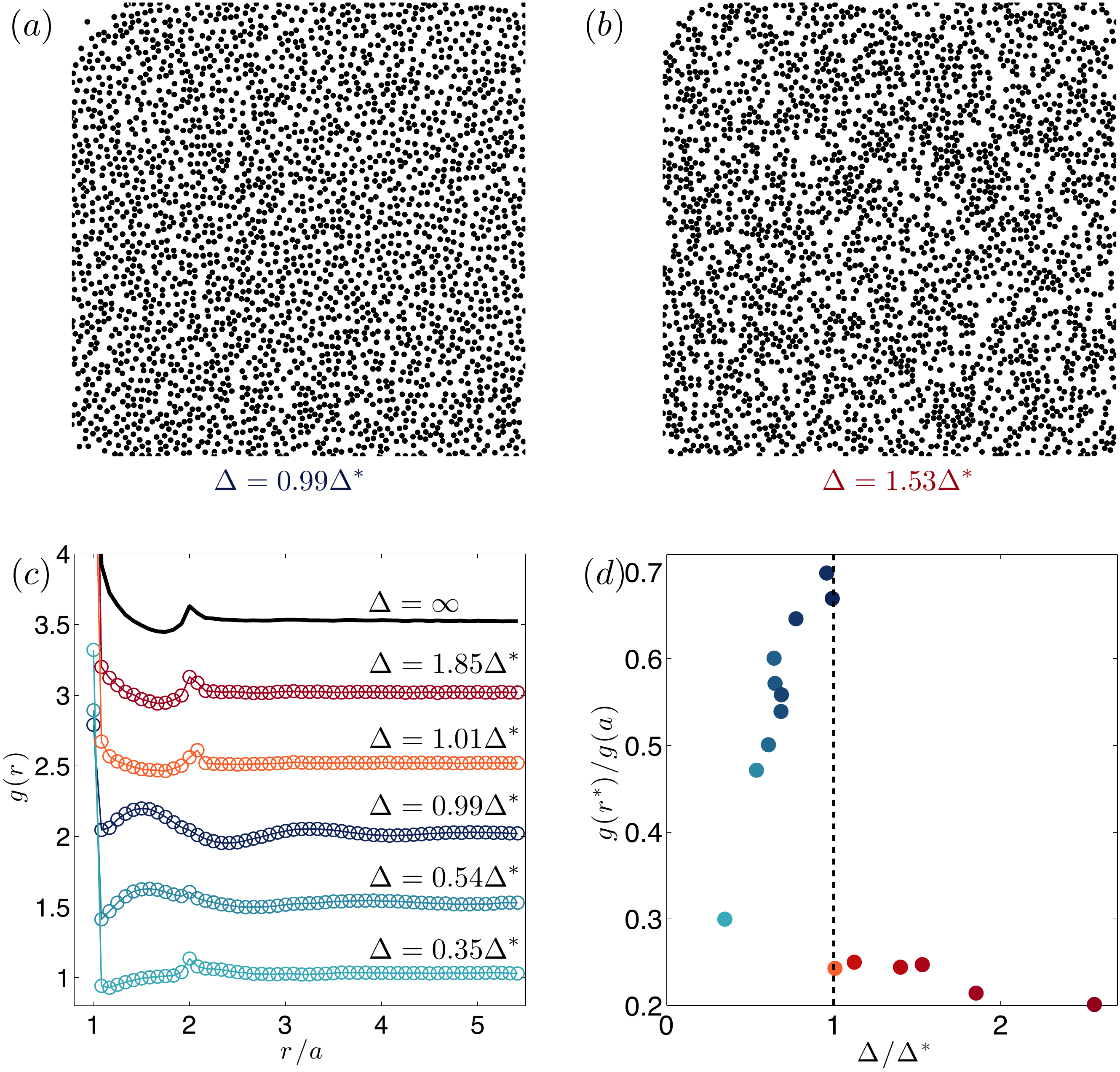}
\end{center}
\caption{(a) Snapshot of an emulsion shaken at $\Delta=0.99\,\Delta^*$ in the reversible regime, $\phi=0.37$. (b) Snapshot of an emulsion shaken at $\Delta=1.53\,\Delta^*$ in the irreversible regime, $\phi=0.36$. Note that the area fraction in (a) and (b) are quasi identical. (c) Strobed radial distribution function $g(r)$ for different driving amplitudes $\Delta$. For sake of clarity the curves are shifted vertically by a $0.5$ constant value. A sharp change in the structure occurs at $\Delta^*$. Above $\Delta^*$ the structure of the emulsion does not depend on $\Delta$ any longer and remains identical to the one found in the initial state common to all the experiments (black curve). (d) Ratio of the fraction of droplets separated by a distance $r^*=1.5\, a$, to the fraction of droplets in close contact, $g(r^*)/g(a)$, plotted versus $\Delta/\Delta^*$. A clear discontinuity occurs at $\Delta=\Delta^*$ revealing a first order structural transition.}
\label{structure}
\end{figure*}

\begin{figure*}
\begin{center}
\includegraphics[width=2\columnwidth]{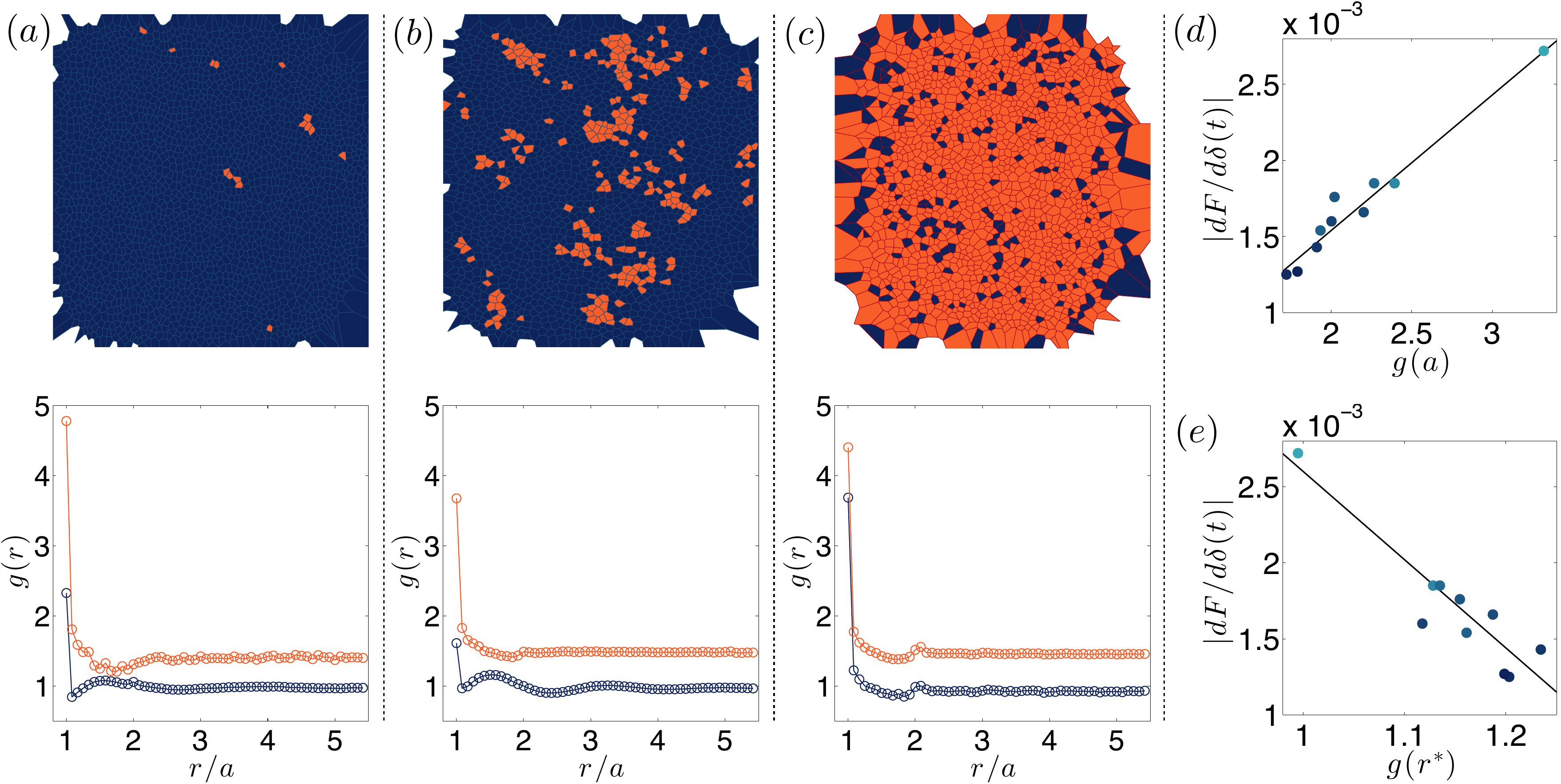}
\end{center}
\caption{(a) Top: Snapshot of the instantaneous  fidelity field for $\Delta=0.54\, \Delta^*$. Each Voronoï cell is colored in blue when associated with a particle that retraces its steps back in the very same cell at the end of the cycle. The cell is colored in orange otherwise. Bottom: Strobed radial distribution function $g(r)$ computed in the reversible (resp. irreversible) cells. Same color code. (b) and  (c) Same plots as in (a) for $\Delta=0.99\, \Delta^*$ and $\Delta=1.53\, \Delta^*$ respectively. The irreversible populations  display a structure akin to a hard-attractive-discs fluid for all the driving amplitudes. Oppositely, the structure in the reversible regions depends on $\Delta$. Below $\Delta^*$, the droplets self-organize to form clusters having a structure similar to a hard-attractive-discs liquid living in a sea of a soft-repulsive-discs liquid. For $\Delta\lesssim\Delta^*$, the correlation length of the translational order is maximal. The typical distance between neighboring particles is maximal as well. Above $\Delta^*$, the instantaneous structure is homogeneous. Only a minute fraction of the droplets return stochastically in their initial Voronoï cell. (d) Slope of the linear part of the instantaneous fidelity decay $|{\rm d} F_\Delta(T,T+t)/{\rm d}\delta (t)|$, plotted as a function of the fraction of particles in close contact $g(a)$ (see Fig.~2b). (e) Slope of the linear part of the instantaneous fidelity decay $|{\rm d} F_\Delta(T,T+t)/{\rm d}\delta (t)|$, plotted as a function of the fraction of particles at a distance $r^*$, $g(r^*)$. Same color code as in all the previous plots.}
\label{phase sep}
\end{figure*}

To gain a deeper physical insight in the collective nature of this discontinuous phase transition, we quantify how much the system differs from its initial conformation in the course of a cycle. We introduce an "instantaneous" fidelity function $F_\Delta(T,T+t)$  where  $t\in[0,T]$, which is defined as the fraction of droplets that occupy the same  Voronoï cell at time $t$ as at the beginning of a cycle. (The positions of the droplets are measured in the frame of  the center of mass of the $N$ droplets to discard  trivial advection.)
We show the variations of $F_\Delta(T,T+t)$ averaged over 15 cycles as a function of the  instantaneous mean displacement $\delta(t)$ in Fig.~2a where the color codes for the driving amplitude.

 For $\Delta<\Delta^*$ (blue curves), and  $\delta(t)/\Delta^*\ll1$,  almost no particle escapes its initial Voronoï cell, and $F_\Delta(T,T+t)$ hardly deviates from 1. For higher  displacements,  $F_\Delta(T,T+t)$    decays linearly with $\delta(t)$ until $t=T/2$. Then starts the echo process, the direction of the flow is reversed and the system should retrace its steps back. As expected, $F_\Delta(T,T+t)$ increases,  linearly again,  as $\delta(t)$ goes back to zero. However, in agreement with Fig.~1f, the dynamics is not fully reversible and the final value $F_\Delta(T,2T)={\cal F}(\Delta)$ is slightly smaller than 1, see Fig.~2b. Looking now at the variations of the instantaneous fidelity with $\delta(t)/\Delta$ in Fig.~2b, we readily note that the  decay rate of the fidelity depends on the driving amplitude, and decreases as  $\Delta^{-1}$. From this observation we infer a pivotal result:  the  conformation of the droplets, at the beginning of a cycle, is a function of the overall driving amplitude. In other words the emulsion self-organizes when periodically driven.

 We now consider the case of the higher amplitudes. For $\Delta>\Delta^*$ (red curves in Fig.~2a), the variations of $F_\Delta(T,T+t)$ are markedly different. During the first half of the cycle, $F_\Delta(T,T+t)$  hardly depends on the driving amplitude: the fidelity decays rapidly  and almost all the droplets rearrange. 
 At the onset of flow reversal some of the rearrangements  are suppressed  by the reversible dynamics and the fidelity increases. However, approaching the end of the cycle the droplets no longer manage to retrace their steps back anymore, and the fidelity decays again to reach a small ${\cal F}(\Delta)$ value. Importantly, in this limit, the system self-organizes into a state that does {\em not} depend on the driving amplitude anymore.

 Self-organization has been already demonstrated in a minimal model for periodically sheared suspensions introduced by Corté \textit{et al}~\cite{corte2008}. This model rely only on two-body contact  interactions to account for the transition from a reversible to an irreversible state reported in \cite{Pine2005,corte2008}. In this model, the macroscopic reversibility upon periodic shear was shown to correspond to  the existence of many accessible absorbing states where the particles do not interact at all in the course of the cycles. Here, we have  described  another possible scenario. Even for $\Delta<\Delta^*$ where the  dynamics is reversible, the particles strongly interact. A macroscopic fraction of the droplets rearrange even for the smallest driving amplitudes, Fig. 2a. The instantaneous position of the individual droplets is not trapped inside an absorbing state. The phenomenology that we observe is actually  similar to that found in soft solids close to the onset of plastic flows~\cite{Hebraud1997,Keim2013}.  The existence of reversible "plastic" events is a generic feature of dry grains, concentrated emulsions and suspensions interacting elastically
when periodically driven  below their yield stress~\cite{Slotterback2012,Keim2013,Regev2013}. %{\color{blue} In addition it is very likely  that concentrated non-Brownian suspension display we can note that even if the dynamics related here is very different from that of the model in \cite{corte2008}, rearrangements of this kind during the cycles must certainly arise in the experiments of \cite{Pine2005}.}

We have shown above that: (i) the loss of fidelity  results from a non-equilibrium phase transition, (ii)  this collective phenomenon is associated to the self-organization of the droplets below the reversibility threshold. We now provide an explicit connection between these two findings and shed light on the nature of the self-organization process. %

\subsection*{Geometrically-protected  reversible states}

Looking at two instantaneous snapshots of the droplets below and above $\Delta^*$ (Fig.~3a, and 3b respectively), one clearly notices that the spatial structure of the emulsion strongly depends on the driving amplitude. The droplets fill the channel in a more homogeneous fashion when undergoing reversible trajectories. To go beyond this qualitative observation, we have plotted the variations of the (strobed) radial distribution function $g(r)$ at the beginning of each cycle, for different driving amplitudes in Fig.~3c. We also show the radial distribution corresponding to the initial conditions common to all the experiments, the highest curve in Fig.~3c. In the initial state, $g(r)$ is characterized by a strongly peaked value at  $r=a$, which is characteristic of the steric repulsion between hard discs. Again, the droplets are non deformed by the flow. A second narrow peak at $r=2a$ suggests that the droplets experience  short range attractive interactions. The position of the droplets loses any  correlation above $r=2a$. This  structure is  analogous to that found in confined fluidized beds made of rigid  discs~\cite{Rouyer2000}, which demonstrates that the effective attractive interactions between the droplets do not result from any physico-chemical adhesion force, but solely stem from hydrodynamic interactions.

In the small amplitude regime, $\Delta\ll\Delta^*$, the structure of the suspension is hardly modified. However, as $\Delta$ is further increased,  the  static structure of the emulsion self-organizes: the sharp peaks at $r=a$ and $r=2a$ decrease while another structure emerges. The  radial distribution displays an additional set of marked oscillations revealing a structure analogous to  a soft-sphere liquids at equilibrium, with a first wider peak located at $r^{*}\sim1.5\,a$ regardless of the driving amplitude. As $\Delta$ approaches $\Delta^*$, the spatial correlation of this liquid-like structure  increases up to distances as large as 7 particle diameters, see Supplementary Figure 2. Conversely, above $\Delta^*$ self-organization does not operate anymore, and the emulsion  weakly deviates from its initial conformation.  The abrupt drop of the fidelity  at $\Delta^*$ coincides with a discontinuous structural change, Fig.~3d. This observation, together with the superposition of two  different marked structures below $\Delta^*$ confirm  a first order phase-transition scenario, and suggests that  the system undergoes a dynamical phase separation. We  now provide a clear demonstration of this hypothesis. 

\begin{figure*}
\begin{center}
\includegraphics[width=0.8\textwidth]{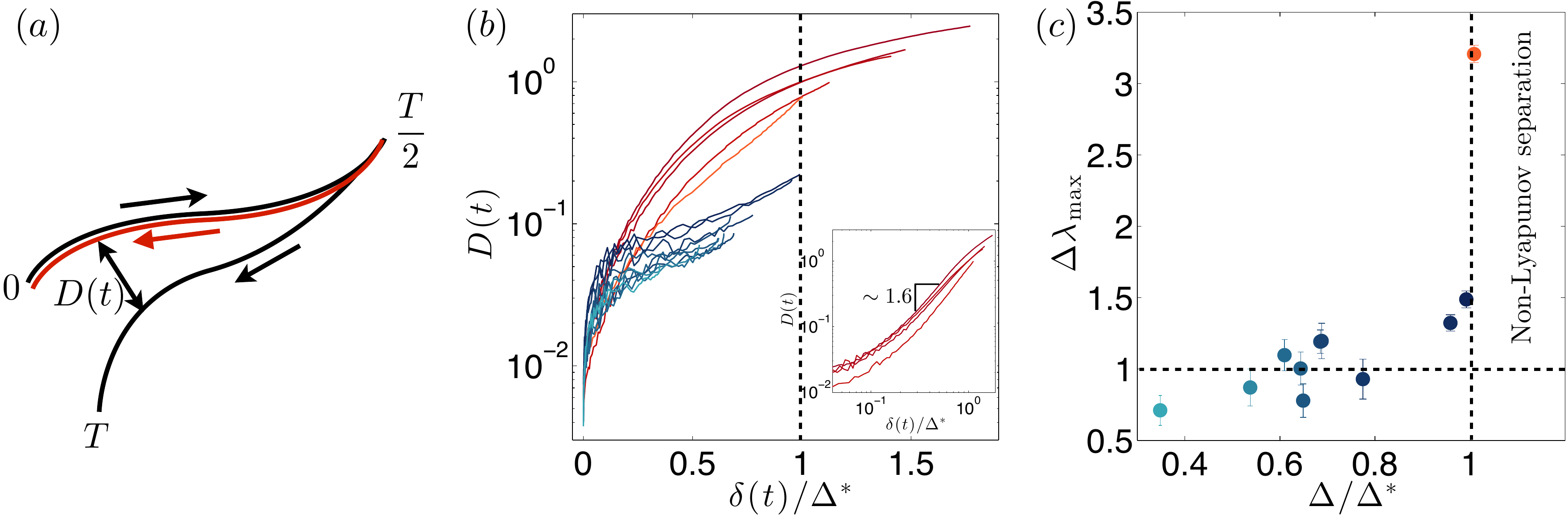}
\end{center}
\caption{(a) Sketch of the trajectory of  the system in its $2N$-dimensional phase space upon periodic driving (black lines). These two paths differ due to the chaotic amplification of any minute non-time reversible perturbation.  Reversing the arrow of time along the first half of the cycle (red curve), these two trajectories are two different realizations of the system dynamics, for initial condition taken at $t=T/2$.  (b) Semi-log plot of $D(t)$ as a function of $\delta(t)/\Delta^*$ for all the driving amplitudes. Same color code as in the previous figures. Inset:  Log-Log plot. Above $\Delta^*$, the separation is no longer exponential but algebraic.  (c) Plot of $\Delta\lambda_{\rm max}$  as a function of $\Delta/\Delta^*$ revealing  memory-loss at the onset of irreversibility.}
\label{lyapunov}
\end{figure*}

In Figs.~4a, b and c we show  instantaneous conformations of the local fidelity field (see Supplementary Movies 4, 5 and 6 for the corresponding strobed movies). Each Voronoï cell, defined at the beginning of the cycle, is colored in blue when associated with a particle that retraces its step back to the very same cell at the end of the cycle.  The cell is colored in orange otherwise. At small amplitudes,  $\Delta\ll\Delta^*$, clusters composed of  very few   irreversible events are dilute in a sea of reversible ones, see Fig.~4a. As $\Delta$ approaches $\Delta^*$,  Fig.~4b, both the typical  size and the number of those irreversible transient clusters increase, thereby inducing the fidelity decay observed in Fig.~1f. The relation between the fidelity loss and the structural properties of the emulsion can now be clearly established. Using these fidelity maps, we plot in Figs.~4a and 4b the radial distribution function of the particles living in the reversible (resp. irreversible) region only. As the driving increases, the spatial structure in the irreversible regions remains unchanged. This structure corresponds to a hard-disc gas with attractive interactions as in the initial (homogeneous) state of the experiment. In contrast, in the reversible regions the droplets self-organize to form a soft-repulsive-disc structure.  At the onset of the fidelity loss, the extent of  translational order is maximal, thereby maximizing the typical distance between the neighboring particles. The instantaneous conformation of the suspension is biphasic: two fluids having different structures coexist in two separated phases. This picture no longer holds above $\Delta^*$, Fig.~4c. All the particles then contribute in the same fashion to the average structure. The emulsion retains its initial spatial ordering, and the strong agitation impedes  phase separation. Most of the particles undergo irreversible trajectories, and only a minute fraction of the Voronoï cells are associated with reciprocal motion. %In this regime cells with a fidelity equal to 1 only pop randomly in the system. 

In order to further emphasize the relation between the dynamical and the structural properties of the emulsion,  we now show that the number of particles that escape their initial Voronoï cell in the course of a cycle is  set by the typical distance between the neighboring droplets. To do so, we plot  the decay  of the instantaneous fidelity per displacement unit: $|{\rm d} F_\Delta(T,T+t)/{\rm d}\delta (t)|$ (slope of the straight lines  in Fig.~2a) as a function of $g(a)$ and $g(r^*)$, see Figs.~4d and 4e. When self-organization occurs, $\Delta<\Delta^*$, the fidelity-decay rate increases linearly with $g(a)$ which is the fraction of droplets in close contact. In contrast, this decay rate decreases with $g(r^{*})$, viz. as the fraction of droplets separated by a finite distance $r^*\sim1.5\,a$ increases. 
In other words, pushing the droplets apart the escape rate from the initial Voronoï cells is restrained: macroscopic reversibility is geometrically protected by a structural self-organization.
\subsection*{Irreversibility and memory loss}

We have hitherto investigated the dynamics of the emulsion in real space from a condensed matter perspective. To gain more physical insight on the origin of reversibility loss upon periodic driving, we now sketch the trajectory of the system in its $2N$-dimensional phase space, black line in Fig.~5a. If the particles were to interact uniquely via (time-reversible) hydrodynamic interactions, this trajectory would be reciprocal. However, in an experiment, or in a numerical simulation,  any minute non-time reversible perturbation along the trajectory is expected to be amplified due to the  chaotic nature of this many-body system~\cite{Aref1984,Pine2005,Drazer2002}.  As a result, the  phase-space trajectory should separate into two different back and forth paths. 
%, this many-body system has to be chaotic~\cite{Aref1984,Pine2005,Drazer2002}
%However, this many-body system has to be chaotic~\cite{Aref1984,Pine2005,Drazer2002}, therefore any minute non-time reversible perturbation along the trajectory is strongly amplified. As a result, the  trajectory separates into two different back, and forth paths. 
Exploiting the invariance upon time reparametrization of the hydrodynamic interaction we can reverse the arrow of time along the first part of the cycle (red curve in Fig.~5a). By doing so,  the two half-trajectories are analogous to two different realizations of the system dynamics for initial conditions taken at $t=T/2$. The asymptotic separation between these two perturbed trajectories after a time  $t$ is measured by computing the  metric distance in  phase space between the conformations of the particles at times $T/2-t$ and $T/2+t$. We denote this distance by $D(t)$. In Fig.~5b $D(t)$ is plotted as a function of the instantaneous mean displacement $\delta(t)$ for all  driving amplitudes $\Delta$. For $\Delta<\Delta^*$,  $D(t)$ displays a clear exponential increase, which allows for measuring the largest Lyapunov exponent of the underlying chaotic dynamics. Above $\Delta^*$, the Lyapunov picture breakdown, and the separation between the trajectories  increases algebraically, see Fig.~5b inset. 

Following~\cite{During2009}, we expect  reversibility to be lost when the chaotic amplification of the microscopic fluctuations causes the system to forget about its initial state in a time smaller than $T/2$, i.e. over a mean displacement smaller than $\Delta$. In more quantitative terms, we expect the fidelity to decay for $\lambda_{\max}\Delta\sim1$, where the largest Lyapunov exponent $\lambda_{\max}$ is defined as $D(t)\sim \exp[\lambda_{\max} \delta(t)]$ asymptotically.  This hypothesis is confirmed by plotting the product $\lambda_{\max}\Delta  $ as a function of $\Delta/\Delta^*$ in Fig.~5c. Below the onset of fidelity decay, the structural self-organization induces variations of $\lambda_{\max}$ that almost exactly balance the increase in the drive amplitude: $\lambda_{\max}\Delta\sim 1$. Conversely, as $\Delta$ exceeds  $\Delta^*$, the path memory is lost, $\lambda_{\max}\Delta>1$ and the fidelity decays accordingly, Fig.~5c.

As a last physical comment, we note that the scaling $\lambda_{\max}\sim 1/\Delta$ is  the signature of the convergence toward an absorbing dynamics. Repeating echo protocols, the system evolves toward  a phase-space region  where $\lambda_{\max}\Delta\sim1$ and where the trajectories are fully reversible (${\cal F}\sim 1$). Hence  the strobed dynamics is frozen, and the system cannot escape this absorbing region. Note however, that these absorbing states  refer uniquely to the strobed dynamics, as opposed to the scenario put forward in~\cite{corte2008,Menon2009} where the dynamics is frozen  along the entire trajectories. Again this situation is very different from that reported here.
We can therefore conjecture that the transition toward irreversibility occurs when all the regions of phase space where $\lambda_{\rm max}$ is small enough  are either out of reach starting from a strongly shaken state, or do not exist.  The latter scenario is consistent with the long-range nature of the hydrodynamic interactions ($\sim1/r^2$ decay in 2D~\cite{Desreumaux}) which strongly couples the entire population of particles regardless of their conformation. %

\section*{Discussion}
 Taking advantage of a quantitative microfluidic experiment, we have demonstrated that periodic shaking results in the emergence of structural order in an ensemble of particles interacting in a time-reversible fashion. Unlike  equilibrium systems,  this geometric self-organization protects the macroscopic reversibility  of the system and realizes a (nonequilibrium) Loschmidt-echo  experiment.  Even though our primary motivation was to gain a deeper insight into the collective dynamics of periodically driven systems,  we anticipate this study to  be of practical relevance. Hydrodynamic interactions in fine-tuned flows have recently been shown to be a potential means to program directed colloidal assembly~\cite{Schneider2011}. Here we have demonstrated that even in the simplest possible flow field (a plug flow),   hydrodynamic interactions can be put to work to induce spatial ordering in an emulsion. Over the last five years, much effort has  been devoted to  synthesizing  non-spherical colloids in the 1-100 microns range, see e.g.~\cite{Yi2013,Dendukuri2009}. Given our observations, using simple AC hydrodynamic actuation could be a powerful tool to trigger and tailor their self-assembly over massive scales, which until now has remained a major practical issue. %A collective effect to engineer self-assembled material is indeed expected to depend only weakly on the   variations in  particle shape and/or physical properties at the single particle level, thereby making this approach  robust and therefore well suited to large-scale production.

\newpage
\section*{Methods}
\subsection*{Microfluidics}
The microfluidic device is a double etched fused silica/quartz  chip  (Micronit Microfluidics) sketched in Supplementary Figure 1. It consists of: (i) a conventional flow-focusing junction used to form monodisperse droplets, (ii)  a dilution module made of an additional  cross junction, and (ii)  two 0.5 cm$\times5$ cm channels.  One of the two wide channels is used to transfer the polydisperse droplets produced in the transient regime of the drop emitter. The other one is used to perform the experimental observations, see e.g. Fig. 1a, and Supplementary Movie 1. The height of the flow-focusing and of the dilution modules is $11\,\mu {\rm m}$ while the two large channels  are $27\,\mu {\rm m}$ high. The fluids are injected using three syringe pumps (Cetoni, Nemesys) connected to the three inlets of the device (inlets (1), (2) and (3) in Supplementary Figure S1). Hexadecane is injected through (1) and the continuous phase (0.1 wt$\%$ SDS (Sigma), 0.2 wt$\%$ fluorescein (Sigma) in deionized water ) is injected both through inlets (2) and (3). The channel surfaces were cleaned in a UV/Ozone cleaner through the quartz surface prior to the experiments. As a result, both the glass and the quartz surfaces are highly hydrophilic, thereby preventing  partial wetting of the walls by the dispersed phase).

The area fraction $\Phi= 0.36$ is obtained by setting the flow rates as $Q_{(1)}=0.040\, \mu {\rm L.s}^{-1}$, $Q_{(2)}=0.140\, \mu {\rm L.s}^{-1}$ and $Q_{(3)}=3 \, \mu {\rm L.min}^{-1}$. The emulsion is periodically driven by imposing an oscillatory flow rate at the outlet (4). To impose the various values of $\Delta$, the magnitude, and period of the sinusoidal oscillations are set to values comprised between $0.02<Q<0.039\, \mu {\rm L.s}^{-1}$ and  $10<T<30\,{\rm s}$ respectively. As expected, the variations of $\Delta$ are linear in $Q$ and $T$.

Prior to any experiment  the system is prepared by imposing a high amplitude sinusoidal oscillation ($Q=0.1\, \mu {\rm L.s}^{-1}$) during several cycles. This protocol is chosen to set  reproducible initial conditions. In order to make sure that the system is in a steady state, the  image acquisition is performed after $1000$ cycles  at  the desired flow rate and period,and  we systematically check that all the measured quantities are statistically stationary. The droplet positions are recorded over  $150$ ($\Delta>\Delta^*$) or $700$ cycles ($\Delta<\Delta^*$).

\subsection*{Imaging and particle tracking}
The device is mounted on a Nikon AZ100 upright macroscope. A Basler Aviator av2300-25gm (4 Megapixel, 8bit) camera is used to record the movement of the droplets in a field of view of  $2.70{\rm mm}\times 2.03{\rm mm}$ at the center of the main channel. The frame rate is set at $24\,{\rm Hz}$. Image acquisition is performed using a custom direct-to-disk Labview code. The particles are detected to a one-pixel accuracy. The particle velocities 
are then computed using the Matlab version of the tracking software developed by  Grier,  Crocker and  Weeks~\cite{Crocker1996}.

%%%
%\bibliographystyle{naturemag}
%\bibliography{biblioreversible.bib}

\begin{thebibliography}{10}
\expandafter\ifx\csname url\endcsname\relax
  \def\url#1{\texttt{#1}}\fi
\expandafter\ifx\csname urlprefix\endcsname\relax\def\urlprefix{URL }\fi
\providecommand{\bibinfo}[2]{#2}
\providecommand{\eprint}[2][]{\url{#2}}

\bibitem{Gorin2006}
\bibinfo{author}{Gorin, T.}, \bibinfo{author}{Prosen, T.},
  \bibinfo{author}{Seligman, T.~H.} \& \bibinfo{author}{\v{Z}nidari\v{c}, M.}
\newblock \bibinfo{title}{{Dynamics of Loschmidt echoes and fidelity decay}}.
\newblock \emph{\bibinfo{journal}{Physics Reports}}
  \textbf{\bibinfo{volume}{435}}, \bibinfo{pages}{33--156}
  (\bibinfo{year}{2006}).

\bibitem{Okuma2011}
\bibinfo{author}{Okuma, S.}, \bibinfo{author}{Tsugawa, Y.} \&
  \bibinfo{author}{Motohashi, A.}
\newblock \bibinfo{title}{{Transition from reversible to irreversible flow:
  Absorbing and depinning transitions in a sheared-vortex system}}.
\newblock \emph{\bibinfo{journal}{Physical Review B}}
  \textbf{\bibinfo{volume}{83}}, \bibinfo{pages}{012503}
  (\bibinfo{year}{2011}).

\bibitem{Aref1984}
\bibinfo{author}{Aref, H.}
\newblock \bibinfo{title}{{Stirring by chaotic advection}}.
\newblock \emph{\bibinfo{journal}{Journal of Fluid Mechanics}}
  \textbf{\bibinfo{volume}{143}}, \bibinfo{pages}{1--21}
  (\bibinfo{year}{1984}).

\bibitem{Slotterback2012}
\bibinfo{author}{Slotterback, S.} \emph{et~al.}
\newblock \bibinfo{title}{{Onset of irreversibility in cyclic shear of granular
  packings}}.
\newblock \emph{\bibinfo{journal}{Physical Review E}}
  \textbf{\bibinfo{volume}{85}}, \bibinfo{pages}{021309}
  (\bibinfo{year}{2012}).

\bibitem{Hebraud1997}
\bibinfo{author}{H\'{e}braud, P.}, \bibinfo{author}{Lequeux, F.},
  \bibinfo{author}{Munch, J.} \& \bibinfo{author}{Pine, D.}
\newblock \bibinfo{title}{{Yielding and Rearrangements in Disordered
  Emulsions}}.
\newblock \emph{\bibinfo{journal}{Physical Review Letters}}
  \textbf{\bibinfo{volume}{78}}, \bibinfo{pages}{4657--4660}
  (\bibinfo{year}{1997}).

\bibitem{Pine2005}
\bibinfo{author}{Pine, D.~J.}, \bibinfo{author}{Gollub, J.~P.},
  \bibinfo{author}{Brady, J.~F.} \& \bibinfo{author}{Leshansky, A.~M.}
\newblock \bibinfo{title}{{Chaos and threshold for irreversibility in sheared
  suspensions}}.
\newblock \emph{\bibinfo{journal}{Nature}} \textbf{\bibinfo{volume}{438}},
  \bibinfo{pages}{997--1000} (\bibinfo{year}{2005}).

\bibitem{Metzger2012}
\bibinfo{author}{Metzger, Bloen}  \& \bibinfo{author}{Butler, Jason E.}
\newblock \bibinfo{title}{{Clouds of particles in a periodic shear flow}}.
\newblock \emph{\bibinfo{journal}{Physics of Fluids}} \textbf{\bibinfo{volume}{24}},
  \bibinfo{pages}{021703} (\bibinfo{year}{2012}).

\bibitem{Keim2013}
\bibinfo{author}{Keim, N.~C.} \& \bibinfo{author}{Arratia, P.~E.}
\newblock \bibinfo{title}{{Yielding and microstructure in a 2D jammed material
  under shear deformation}}.
\newblock \emph{\bibinfo{journal}{Soft Matter}} \textbf{\bibinfo{volume}{9}},
  \bibinfo{pages}{6222-6225} (\bibinfo{year}{2013}).

\bibitem{swendsen2008}
\bibinfo{author}{Swendsen, R.~H.}
\newblock \bibinfo{title}{{Explaining irreversibility}}.
\newblock \emph{\bibinfo{journal}{American Journal of Physics}}
  \textbf{\bibinfo{volume}{76}}, \bibinfo{pages}{643} (\bibinfo{year}{2008}).

\bibitem{corte2008}
\bibinfo{author}{Cort\'{e}, L.}, \bibinfo{author}{Chaikin, P.~M.},
  \bibinfo{author}{Gollub, J.~P.} \& \bibinfo{author}{Pine, D.~J.}
\newblock \bibinfo{title}{{Random organization in periodically driven
  systems}}.
\newblock \emph{\bibinfo{journal}{Nature Physics}}
  \textbf{\bibinfo{volume}{4}}, \bibinfo{pages}{420--424}
  (\bibinfo{year}{2008}).

\bibitem{Metzger2010a}
\bibinfo{author}{Metzger, B.} \& \bibinfo{author}{Butler, J.}
\newblock \bibinfo{title}{{Irreversibility and chaos: Role of long-range
  hydrodynamic interactions in sheared suspensions}}.
\newblock \emph{\bibinfo{journal}{Physical Review E}}
  \textbf{\bibinfo{volume}{82}}, \bibinfo{pages}{1--4} (\bibinfo{year}{2010}).

\bibitem{Mangan2008}
\bibinfo{author}{Mangan, N.} \& \bibinfo{author}{Reichhardt, C.}
\newblock \bibinfo{title}{{Reversible to Irreversible Flow Transition in
  Periodically Driven Vortices}}.
\newblock \emph{\bibinfo{journal}{Physical Review Letters}}
  \textbf{\bibinfo{volume}{100}}, \bibinfo{pages}{187002}
  (\bibinfo{year}{2008}).

\bibitem{Menon2009}
\bibinfo{author}{Menon, G.~I.} \& \bibinfo{author}{Ramaswamy, S.}
\newblock \bibinfo{title}{{Universality class of the reversible-irreversible
  transition in sheared suspensions}}.
\newblock \emph{\bibinfo{journal}{Physical Review E}}
  \textbf{\bibinfo{volume}{79}}, \bibinfo{pages}{1--4} (\bibinfo{year}{2009}).

\bibitem{Desreumaux}
\bibinfo{author}{Desreumaux, N.}, \bibinfo{author}{Caussin, J.-B.},
  \bibinfo{author}{Jeanneret, R.}, \bibinfo{author}{Lauga, E.} \&
  \bibinfo{author}{Bartolo, D.}
\newblock \bibinfo{title}{{Hydrodynamic fluctuations in confined particle-laden
  fluids}}.
\newblock \emph{\bibinfo{journal}{Physical Review Letters}} \textbf{\bibinfo{volume}{111}}, \bibinfo{pages}{118301} (\bibinfo{year}{2013}).

\bibitem{Abate2007a}
\bibinfo{author}{Abate, A.} \& \bibinfo{author}{Durian, D.}
\newblock \bibinfo{title}{{Topological persistence and dynamical
  heterogeneities near jamming}}.
\newblock \emph{\bibinfo{journal}{Physical Review E}}
  \textbf{\bibinfo{volume}{76}}, \bibinfo{pages}{021306}
  (\bibinfo{year}{2007}).

\bibitem{Dauchot2010a}
\bibinfo{author}{Dauchot, O.}, \bibinfo{author}{Durian, D.~J.} \&
  \bibinfo{author}{van Hecke, M.}
\newblock \emph{\bibinfo{title}{Dynamical Heterogeneities in Glasses, Colloids,
  and Granular Media}}, chap. \bibinfo{chapter}{Dynamical heterogeneities in
  grains and foams} (\bibinfo{publisher}{Oxford University Press},
  \bibinfo{address}{Oxford}, \bibinfo{year}{2011}).

\bibitem{guyon2001}
\bibinfo{author}{Guyon, {\'E}.}, \bibinfo{author}{Hulin, J.} \&
  \bibinfo{author}{Petit, L.}
\newblock \emph{\bibinfo{title}{Physical Hydrodynamics}}
  (\bibinfo{publisher}{Oxford University Press}, \bibinfo{address}{Oxford},
  \bibinfo{year}{2001}).

\bibitem{Regev2013}
\bibinfo{author}{Regev, Ido}, \bibinfo{author}{Lookman, Turab} \& \bibinfo{author}{Reichhardt, Charles}
\newblock \bibinfo{title}{{Onset of irreversibility and chaos in amorphous solids under periodic shear}}.
\newblock \emph{\bibinfo{journal}{Physical Review E}}
  \textbf{\bibinfo{volume}{88}}, \bibinfo{pages}{062401} (\bibinfo{year}{2013}).

\bibitem{Rouyer2000}
\bibinfo{author}{Rouyer, F.}, \bibinfo{author}{Lhuillier, D.},
  \bibinfo{author}{Martin, J.} \& \bibinfo{author}{Salin, D.}
\newblock \bibinfo{title}{{Structure, density, and velocity fluctuations in
  quasi-two-dimensional non-Brownian suspensions of spheres}}.
\newblock \emph{\bibinfo{journal}{Physics of Fluids}}
  \textbf{\bibinfo{volume}{12}}, \bibinfo{pages}{958} (\bibinfo{year}{2000}).

\bibitem{Drazer2002}
\bibinfo{author}{Drazer, G.}, \bibinfo{author}{Koplik, J.},
  \bibinfo{author}{Khusid, B.} \& \bibinfo{author}{Acrivos, A.}
\newblock \bibinfo{title}{{Deterministic and stochastic behaviour of
  non-Brownian spheres in sheared suspensions}}.
\newblock \emph{\bibinfo{journal}{Journal of Fluid Mechanics}}
  \textbf{\bibinfo{volume}{460}}, \bibinfo{pages}{307--335}
  (\bibinfo{year}{2002}).

\bibitem{During2009}
\bibinfo{author}{D\"{u}ring, G.}, \bibinfo{author}{Bartolo, D.} \&
  \bibinfo{author}{Kurchan, J.}
\newblock \bibinfo{title}{{Irreversibility and self-organization in
  hydrodynamic echo experiments}}.
\newblock \emph{\bibinfo{journal}{Physical Review E}}
  \textbf{\bibinfo{volume}{79}}, \bibinfo{pages}{1--4} (\bibinfo{year}{2009}).

%\bibitem{Frenkel2008}
%\bibinfo{author}{Frenkel, D.}
%\newblock \bibinfo{title}{{Random organization: Ordered chaos}}.
%\newblock \emph{\bibinfo{journal}{Nature physics}}
%  \textbf{\bibinfo{volume}{4}}, \bibinfo{pages}{345--346} (\bibinfo{year}{2008}).

\bibitem{Schneider2011}
\bibinfo{author}{Schneider, T.~M.}, \bibinfo{author}{Mandre, S.} \&
  \bibinfo{author}{Brenner, M.~P.}
\newblock \bibinfo{title}{{Algorithm for a Microfluidic Assembly Line}}.
\newblock \emph{\bibinfo{journal}{Physical Review Letters}}
  \textbf{\bibinfo{volume}{106}}, \bibinfo{pages}{094503}
  (\bibinfo{year}{2011}).

\bibitem{Yi2013}
\bibinfo{author}{Yi, G.-R.}, \bibinfo{author}{Pine, D.~J.} \&
  \bibinfo{author}{Sacanna, S.}
\newblock \bibinfo{title}{{Recent progress on patchy colloids and their
  self-assembly}}.
\newblock \emph{\bibinfo{journal}{Journal of physics. Condensed matter : an
  Institute of Physics journal}} \textbf{\bibinfo{volume}{25}},
  \bibinfo{pages}{193101} (\bibinfo{year}{2013}).

\bibitem{Dendukuri2009}
\bibinfo{author}{Dendukuri, D.} \& \bibinfo{author}{Doyle, P.~S.}
\newblock \bibinfo{title}{{The Synthesis and Assembly of Polymeric
  Microparticles Using Microfluidics}}.
\newblock \emph{\bibinfo{journal}{Advanced Materials}}
  \textbf{\bibinfo{volume}{21}}, \bibinfo{pages}{4071--4086}
  (\bibinfo{year}{2009}).

\bibitem{Crocker1996}
\bibinfo{author}{Crocker, J.} \& \bibinfo{author}{Grier, D.}
\newblock \bibinfo{title}{{Methods of digital video microscopy for colloidal
  studies}}.
\newblock \emph{\bibinfo{journal}{Journal of colloid and interface science}}
  \textbf{\bibinfo{volume}{179}}, \bibinfo{pages}{298--310}
  (\bibinfo{year}{1996}).

\end{thebibliography}

\section*{Acknowledgments}
 We acknowledge illuminating discussions with Gustavo Düring and Jorge Kurchan about the memory-loss picture. We thank Bertrand Levaché for help with the experiments. We acknowledge Laurette Tuckerman for much useful comments.

\section*{Author contributions}
R. J. carried out the experiments and processed the data. D. B. and R. J. discussed and interpreted results. D. B. and R. J. wrote the manuscript.

\section*{Additional information}
Supplementary Information accompanies this paper.

\includepdf[pages={{},1,{},2}]{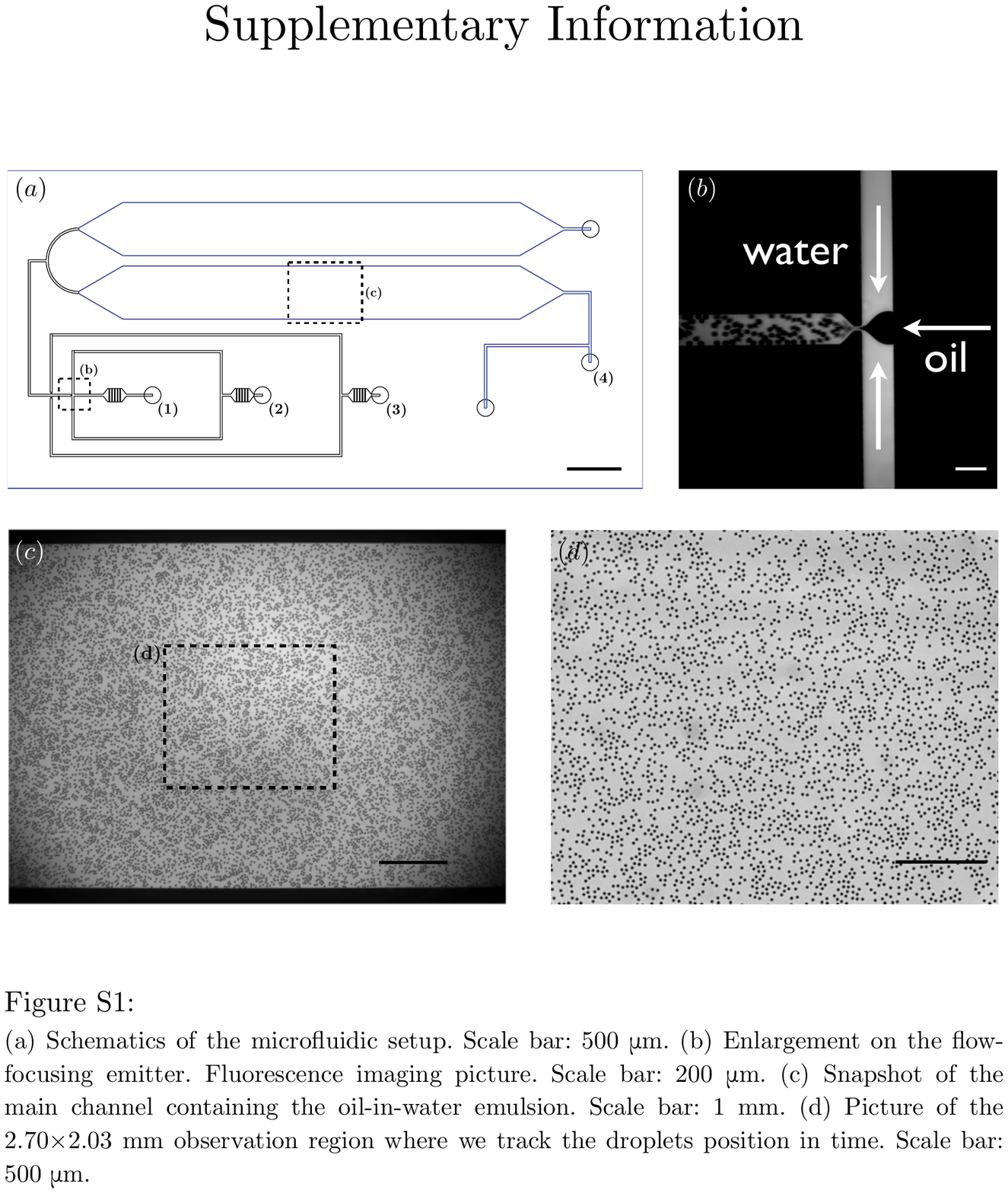}
\includepdf[pages={{},-}]{supp.pdf}
\end{document}